\documentclass[manuscript]{aastex}
\begin{document}
\title{On The Accretion Rates of SW Sextantis Nova-Like
Variables}

\author{Ronald-Louis Ballouz}
\affil{Dept. of Astronomy and Astrophysics,
Villanova University,
Villanova, PA 19085,
email: ronald-louis.ballouz@villanova.edu}

\author{Edward M. Sion}
\affil{Dept. of Astronomy \& Astrophysics,
Villanova University,
Villanova, PA 19085, 
e-mail: edward.sion@villanova.edu}

\begin{abstract}
We present accretion rates for selected samples of nova-like variables having IUE archival spectra
and distances uniformly determined using an infrared method by Knigge (2006). A comparison with accretion rates derived 
independently with a multi-parametric optimization modeling approach by Puebla et al.(2007) is carried out.  
The accretion rates of SW Sextantis nova-like systems are compared with the accretion rates of non-SW Sextantis
systems in the Puebla et al. sample and in our sample, which was selected in the orbital period range of three to four and a half hours,
with all systems having distances using the method of Knigge (2006). Based upon the two independent modeling approaches, we find no significant
difference between the accretion rates of SW Sextantis systems and non-SW Sextantis nova-like systems insofar as optically thick disk models 
are appropriate. We find little evidence to suggest that the SW Sex stars have higher accretion rates than other nova-like CVs above the period
gap within the same range of orbital periods. 
\end{abstract}

Subject Headings: Stars: cataclysmic variables, white dwarfs, Physical
Processes: accretion, accretion disks

\section{Introduction}

Cataclysmic variables (CVs) are short-period binaries in which a
late-type, Roche-lobe-filling main- sequence dwarf transfers gas through
an accretion disk onto a rotating, accretion-heated white dwarf (WD). The
nova-like variables are a non-magnetic subclass of CVs in which the mass-transfer rate
tends to be high and the light of the system is typically dominated by a very bright
accretion disk (Warner 1995). The spectra of nova-like variables resemble those of
classical novae (CNe) that have settled back to quiescence. However,
nova-like variables have never had a recorded CN outburst or any outburst. Hence their
evolutionary status remains unknown. They could be close to having their next
CN explosion, or they may have had an unrecorded explosion, in the recent past. 
Their distribution of orbital periods reveals a large concentration of systems in the 
range between three and four hours, the former period being the upper boundary of the CV period gap
where very few CVs are found. Some nova-likes 
(classified as the VY Sculptoris systems) show the behavior of being in a high optical
brightness state for most of the time, but then, for no apparent reason, plummeting into a deep 
low optical brightness state with little or no ongoing accretion. Then, just as unpredictably,
their optical brightness returns to the high state(cf. Honeycutt \& Kafka 2004 and references therein). 
These precipitous drops in brightness are possibly related to the cessation 
of mass transfer from the K-M dwarf secondary star either by starspots that drift into position 
under the inner Lagrangian point, L1 (Livio \& Pringle 1998) or irradiation feedback in which an 
inflated outer disk can modulate the mass transfer from the secondary by blocking its irradiation 
by the hot inner accretion disk region (Wu et al. 1995). Other nova-like systems, the UX UMa subclass,
do not appear to exhibit low states but remain in a state of high accretion, sometimes 
referred to as dwarf novae stuck in permanent outburst. It is widely assumed that the absence of dwarf 
novae outbursts in nova-likes is explained by their mass transfer rates being above a critical 
threshold where the accretion rates are so high that the accretion disk is largely ionized thus 
suppressing the viscous-thermal instability (the disk instability mechanism or DIM) which drives 
dwarf nova limit cycles (Shafter, Cannizzo and Wheeler 1986).

Until recently, the accretion rates of nova-likes, including SW Sex stars, have been reported for only a few
individual systems from a variety of model analyses of their optical, FUV spectra or X-ray spectra.
Optical determinations of the accretion rates in nova-likes are based upon estimates of their disk luminosity using distance
estimates or clues (Patterson 1984).
The absolute magnitudes of the accretion disks in nova-likes reveal that their accretion rates are similar to those derived for dwarf novae
during their outbursts (Warner 1995). 
Unfortunately, the distances of nova-like variables remain uncertain due to the scarcity of trigonometric parallaxes
and the absence of a reliable usable relation for nova-like variables between their absolute magnitude at maximum light versus orbital period
similar to what exists for dwarf novae.
A more systematic study of a larger number of systems is clearly needed in order to compare accretion rates 
among different subgroups of CVs. One recent statistical study (Puebla, Diaz \& Hubeny 2007), utilizing a 
multi-parametric optimization model fitting method, explored how well current optically thick accretion disk models fit
the FUV spectra of nova-likes and old novae in a sample of 33 nova-like and old novae.
They found the average value of \.{M} for nova-like systems was $\sim 9.3\times 10 ^{-9}$ M$_{\sun}$ yr$^{-1}$. 

Among the nova-like variables is a subclass, the SW Sextantis stars, which display
a multitude of observational characteristics: orbital periods between 3 and 4 hours, 
up to one-half of the known SW Sex systems are non-eclipsing and roughly one-half show deep eclipses of the WD by the secondary, 
thus requiring high inclination angles, single-peaked emission lines despite the high inclination, and high 
excitation spectral features including He II (4686) emission and strong Balmer emission on a blue continuum,
high velocity emission S-waves with maximum blueshift near phase $\sim$  0.5, delay of emission line
radial velocities relative to the motion of the WD, and central absorption dips in the emission lines
around phase $\sim$ 0.4 - 0.7 (Rodriguez-Gil, Schmidtobreick \& Gaensicke 2007; Hoard et al. 2003).
The SW Sex stars appear to be
intrinsically luminous as indicated by the apparent brightnesses of systems like DW UMa despite their being viewed at high orbital inclinations 
of 80 degrees and higher (Rodriguez-Gil et al.2007). A picture of very high secular mass transfer rates is supported by the presence of very hot
white dwarfs in the rare SW Sex systems observed in a low state (e.g. DW UMa's white dwarf). 
The white dwarfs in many, if not all, of these systems are suspected of 
being magnetic (Rodriguez-Gil et al. 2007). However,
the case for magnetic white dwarfs in the SW Sex stars remains highly speculative and does not
consistently account for the spectroscopic and photometric characteristics from system to system (see for example
Hoard et al. 2003). It has also been asserted that the SW Sextantis subclass of nova-like variables have higher accretion rates
than other nova-like systems (Rodriguez-Gil et al. 2007). Indeed, twenty-seven out of thirty-five SW Sex stars listed by Rodriguez et al.(2007)
have orbital periods concentrated in the range of 3 hours to 4.5 hours where nova-like systems tend to accumulate. As asserted in
Rodriguez et al., either the SW Sex stars have "an average mass transfer rate well above that of their CV cousins" or another source of 
luminosity exists. Since these objects are found near the upper boundary of 
the period gap, their study is of critical importance to understanding CV evolution as they 
enter the period gap (Rodriguez-Gil et al.2007). 

In the SW Sex systems (except those observed during low states), the accretion disk flux completely dominates the FUV 
wavelength range. The white dwarf contribution is expected to be minimal in these systems because their
disks are thick and luminous and because at high inclination the inner disk, boundary layer and white dwarf
should be significantly obscured by vertical structure in the disk. Therefore, it is entirely reasonable that
the analysis of a nova-like system in a high state be carried out with optically thick steady state accretion disk models in which
the accretion rates are determined from fitting the continuum slopes and Lyman Alpha profiles with the fits constrained
by the system distance, if known, and by parameters like the inclination angle and white dwarf mass if reliably known.
This is the same model fitting strategy that we employed to determine the accretion rates of 46 dwarf novae in outburst
from disk modeling of their IUE archival FUV spectra (Hamilton et al. 2007).

An important question is whether or not the SW Sex really have higher than average 
accretion rates compared with other nova-like systems as asserted by Rodriguez et al. 2007).
If they do not have higher than average mass transfer rates, then what is the 
source of their higher luminosities? In order to test this assertion, we have determined accretion rates of SW Sex nova-likes with
non-SW Sex nova-likes within approximately the same orbital period range. 
Our primary goal in this work is to examine the accretion rates of the SW Sextantis subclass of nova-like variables

\section{System Parameters and Distances of Nova-Like Variables}

In order to constrain the synthetic spectral fitting and reduce the number
of free parameters, we search the published literature for the most accurately
known system parameters. This included the compilations in Ritter \& Kolb (2003)
and the Goettingen CVCat website as well as publications documented in the SAO/NASA Astrophysics
Data Service (ADS). The most critical parameter for the model fitting, the distance, is the
least known. We conducted an exhaustive search of the literature for previous published distance estimates.
There is only one nova-like, RW Tri, with a reliable trigonometric parallax measurement. 
Unlike dwarf novae where there exists a
correlation between their absolute magnitude at maximum and their orbital period, there is no such 
relation for the nova-likes. However, a new 
method (Knigge 2006) utilizing 2MASS JHK photometry and the observed properties of CV donor stars
has proven useful for constraining nova-like distances. At present, this is the only reliable handle one has on
nova-like distances (Warner 2008). For each system, 
we obtained the J,H,K apparent magnitudes from 2MASS. For a given orbital period, Knigge (2006)
provides absolute J, H and K magnitudes based upon his semi-empirical donor sequence for 
CVs. If it is assumed that the donor provides 100\% of the light in J, H and K, then the 
distance is a strict lower limit. If the donor emits 33\% of the light (the remainder being
accretion light), then an approximate upper limit is obtained. The latter limit is a factor of
1.75 times the lower limit distance. For moderately bright CVs, interstellar reddening is expected to have a 
negligible effect on the IR photometry used to estimate the distances.
The adopted distance ranges which are used as constraints in the 
synthetic spectral fitting procedure
are given below (see Table 1). 

For our comparative study, we selected all SW Sex and non-SW Sex nova-like variables within the orbital period range
of 3 hours to 4.5 hours for which usable IUE archival spectra exist. This period range is where 77\% of the known SW Sex stars are found. 
The SW Sex status was confirmed by a 
comparison of the census in Table 6 of Rodriguez et al. (2007) with the latest census of membership in the SW Sex class
given in Don Hoard's Big List of SW Sex Stars.
\footnote[1]{See the Big List of SW Sextantis Stars at \url{http://spider.ipac.caltech.edu/staff/hoard/biglist.html}.} 

Within the 3 to 4.5 hour period range, the following objects are assigned "Definite" SW Sex membership status: V442 Oph, SW Sex,
AH Men, WX Ari, BP Lyn and UU Aqr. The systems HL Aqr and LN UMa are listed Hoard's Big List as "Probable" members
while being listed in Rodriguez et al. (2007) as bona fide SW Sex stars. We have retained these two objects as SW Sex
members.  

For the non-SW Sex nova-like systems within the same orbital period for which there are usable IUE archival spectra obtained during
their high brightness state, we selected the following systems: LQ Peg, MV Lyr, TT Ari, VZ Scl, BZ Cam, and CM Del. We note that TT Ari
and VZ Scl are listed in Hoard's Big List as "Possible" SW Sex membership status. Therefore, we compute the average accretion rates below, 
with and without their inclusion.

In Table 1, where we list the adopted parameters for the orbital period (hours), 
the apparent V-magnitude, 
the inclination $\it{i}$ ($\degr$), the white dwarf mass (M$_{\sun}$), the interstellar reddening, E($\bv$),
and the distance in parsecs. In our model fitting procedure, these published parameters are used as initial
guesses but if the resulting fits are unsatisfactory, then we allow the relevant parameters to vary in 
the model fitting.

\begin{deluxetable}{lcccccccc}
\tablecaption{Nova-Like System Parameters}
\tablenum{1}
\tablecolumns{8}
\tablehead{
\colhead{System}
&\colhead{subtype}
&\colhead{P$_{orb}$ (hours)}
&\colhead{V}
&\colhead{M$_{wd}$ (M$_{\sun}$)}
&\colhead{\it{i}($\degr$)}
&\colhead{E ($\bv$)}
&\colhead{d$_{knigge}$ (pc)}
&\colhead{d$_{lit}$ (pc)}
}
\startdata
V442 Oph & VY,SW &2.98&12.6& - & - & 0.22 & 153-268 &130\\
SW Sex &UX,SW& 3.24&14.3& 0.5 & $> 75$: & 0.0 &243-426&450\\
AH Men   & SW & 3.05& 13.2  & - &- & 0.12 & 91-160 & 120  \\
HL Aqr   & SW & 3.25 & 13.3 & - & - & 0.05 & 174-304 & 213 \\
WX Ari   & SW  & 3.34 & 15.3 & -  & 72:& - & 258.7-453 & 468 \\
LN UMa   & SW & 3.46 & 14.6  & - & - &  - & 349-610 &  405\\
BP Lyn   & SW & 3.67 & 14.5  & - & - & - & 251-440  & 344\\
UU Aqr   & SW & 3.93 & 13.3  & $0.67\pm0.14$ & $78\pm2$ & - &  174-304 & 208\\
LQ Peg   & NSW & 2.99 & 14.4 & -& -& -& 270-472 & 350\\
MV Lyr   & NSW & 3.18 & 11.8 & $0.73\pm0.10$ & 12  & - & 431-754 & 442 \\
TT Ari   & NSW & 3.30 & 9.5   & -  & - & -&63-109 & 65 \\
VZ Scl   & NSW  & 3.47 & 15.6 & 1: & 90: & - & 490-858 & 566 \\
BZ Cam   & NSW & 3.69 & 12.0 & -  & - & - & 235-412 & 258\\
CM Del   & NSW & 3.89 & 13.4 & $0.48\pm0.15$ & $73\pm47$ & - & 229-401 & 241 \\
VY Scl   & NSW & 5.57: & 12.1 & $1.22\pm0.22$  & $30\pm10$ & - & 196-343  & 337\\
\hline
\enddata
\end{deluxetable}

\section{Far Ultraviolet Spectroscopic Observations}

All the spectral data were obtained from the Multimission Archive at Space Telescope (MAST) 
IUE archive are in a high activity state, very near or at outburst.  We restricted 
our selection to those systems with SWP spectra, with resolution of 5\AA\ and a spectral 
range of 1170\AA\ to 2000\AA.  All spectra were taken through the large aperture at 
low dispersion.  When more than one spectrum with 
adequate signal-to-noise ratio was available, the spectra were co-added or
the two best spectra were analyzed. 
In Table 2, an observing log of the IUE archival spectra is presented in which by 
column: (1) lists the SWP spectrum number, (2)  the aperture diameter, (3) the 
exposure time in seconds, (4) the date and time of the observation, (5)
 the continuum to background counts, and (6) the brightness state of the system.  
Transition refers to an intermediate state between the highest optical brightness state
and the deepest low state.

\begin{deluxetable}{lccccccc}  
 
\tablecaption{IUE Observing Log}
\tablenum{2}
\tablewidth{0pc}
\tablecolumns{8}
\tablehead{
\colhead{System}
&\colhead{subtype}                                   
&\colhead{SWP}
&\colhead{t$_{exp}$}
&\colhead{Disp.}
&\colhead{Ap.}
&\colhead{Date of Obs.}
&\colhead{State}
}
 \startdata        
V442 Oph& SW & 14731 &  6600&   LOW&    Lg&1981-12-8 & Intermed.? \\
SW Sex& SW & 21534+21535+21536   &  2400/2400/3240 &   LOW&   Lg &1983-11-13 & High \\
AH Men   & SW  & 43037    & 2880  &  LOW & Lg   &1991-11-8 & High \\ 
HL Aqr   & SW  & 23325    & 3600 & LOW & Lg & 1984-6-24 & High \\
WX Ari   & SW  &  55953 & 14100    &  LOW & Lg  &1995-9-17   &  High \\    
LN UMa   & SW  & 40948    & 7200& LOW& Lg & 1991-2-28 & High\\
BP Lyn   & SW   & 32940   & 7200& LOW & Lg& 1988-2-18 & High \\
UU Aqr   & SW & 51249 & 3600     &  LOW&  Lg & 1994-6-29 & High \\
LQ Peg   & NSW & 17367& 2400    & LOW& Lg& 1982-7-6 & High \\
MV Lyr   & NSW & 07296 & 5400   & LOW & Lg & 1979-12-2 & High \\ 
TT Ari   & NSW & 42491 & 420    & LOW & Lg& 1991-9-17 & High \\
VZ Scl   & NSW & 23021 & 25500 & LOW & Lg&  1984-5-15 & High \\
BZ Cam   & NSW& 21251& 1800& LOW& Lg& 1983-10-07 & High \\
CM Del   & NSW& 14707 & 3600 & LOW & Lg & 1981-8-10 & High \\
VY Scl   & NSW& 32594   & 2100 & LOW & Lg & 1987-12-23 & High \\
\hline
\enddata
\end{deluxetable}

In the case of those systems not covered by the AAVSO,
their activity state was assessed based upon either mean photometric 
magnitudes taken from the Ritter \& Kolb(2003) catalogue or from IUE 
Fine Error Sensor (FES) measurements at the time of the IUE observation. 
In addition, the presence of P-Cygni profiles, absorption lines, and a 
comparison with spectral data and flux levels of other systems 
during different activity states was used to ascertain the state of the system.
The reddening of the systems was determined based upon all estimates 
listed in the literature. The three principal sources of reddening were
the compilations of Verbunt (1987), laDous (1991) and Bruch \& Engel
(1994). The spectra were de-reddened with the IUERDAF IDL routine UNRED.

\section{Synthetic Spectral Fitting Models}

We adopted model accretion disks from the optically thick disk model grid 
of Wade \& Hubeny (1998). In these accretion disk models, the innermost disk radius, R$_{in}$, is fixed at a fractional white dwarf radius
of $x = R_{in}/R_{wd} = 1.05$. The outermost disk radius, R$_{out}$, was chosen so that T$_{eff}(R_{out})$ is near 10,000K
since disk annuli beyond this point, which are cooler zones with larger radii, would provide only a very small contribution to the 
mid and far UV disk flux, particularly the SWP FUV bandpass. The mass transfer rate is assumed to be the same for all radii.
Thus, the run of disk temperature with radius is taken to be:

\begin{equation}
T_{eff}(r)= T_{s}x^{-3/4} (1 - x^{-1/2})^{1/4}
\end{equation}

where  $x = r/R_{wd}$
and $\sigma T_{s}^{4} =  3 G M_{wd}\dot{M}/8\pi R_{wd}^{3}$

Limb darkening of the disk is fully taken into account in the manner described by Diaz et al. (1996) involving the Eddington-Barbier relation,
the increase of kinetic temperature with depth in the disk, and the wavelength and temperature dependence of the
Planck function. The boundary layer contribution to the model flux is not included. However, 
the boundary layer is expected to contribute primarily in the extreme ultraviolet below the Lyman limit. 

The disk is divided into a set of ring annuli. The vertical structure of each
ring is computed with TLUSDISK (Hubeny
1990), which is a derivative of the stellar atmosphere
program TLUSTY (Hubeny 1988). The spectrum synthesis program SYNSPEC
described by Hubeny \& Lanz (1995) is used to solve
the radiative transfer equation to compute the local, rest
frame spectrum for each ring of the disk. In addition to
detailed proÐles of the H and He lines, the spectrum synthesis
includes metal lines up to Nickel (Z = 28). The accretion disks are computed in LTE and
the chemical composition of the accretion disk is kept fixed at solar
values in our study.

Theoretical, high gravity, photospheric spectra were computed by first using the code
TLUSTY version 200(Hubeny 1988) to calculate the atmospheric structure and SYNSPEC version 48 (Hubeny and Lanz 1995)
to construct synthetic spectra. We compiled a library of photospheric spectra  
covering the temperature range from 15,000K to 70,000K in increments of 1000 K, and a surface gravity range, 
log $g = 7.0 - 9.0$, in increments of 0.2 in log $g$.

After masking emission lines in the spectra, our normal procedure, in general, is to
determine separately for each spectrum, the best-fitting white dwarf-only 
model, the best-fitting accretion disk-only model, the best fitting combination of a white dwarf
plus an accretion disk, and the best-fitting two-temperature white dwarfa (to include an accretion belt or ring).
Using two $\chi^{2}$ minimization routines, either IUEFIT for disks-alone and photospheres-alone or DISKFIT 
for combining disks and photospheres or two-temperature white dwarfs, $\chi^{2}$ values and a scale factor 
were computed for each model or combination of models. 
The scale factor, $S$, normalized to a kiloparsec 
and solar radius, can be related to the white dwarf radius R through:  
$F_{\lambda(obs)} = S H_{\lambda(model)}$, where $S=4\pi R^2 d^{-2}$, and $d$ is the 
distance to the source. For the white dwarf radii, we use the mass-radius relation from the evolutionary 
model grid of Wood (1995) for C-O cores. The best-fitting model or combination of models 
was chosen based not only upon the minimum $\chi^{2}$ value achieved, but the goodness of fit of the 
continuum slope, the goodness of fit to the observed Lyman Alpha region 
and consistency of the scale factor-derived distance with the adopted 
Knigge (2006) distance for each system. orthe

For a non-magnetic nova-like variable during its high state, it is reasonable to expect
that a steady state optically thick accretion disk might provide a successful fit. Therefore, our modeling procedure
is the same as the procedures carried out by Hamilton et al. (2007) for the entire IUE archive of 46 dwarf novae in outburst.
For the nova-like variables during their high brightness states, we first try accretion disk models which
satisfy both the continuum slope and Lyman Alpha line width in a single model.  
We use published parameters like the WD mass and inclination ONLY as an initial guess in searching for the 
best-fitting accretion disk models. If the parameters are published but not considered reliable or if they 
are entirely absent, then for each systems's spectrum, we carry out fits for every combination of \.{M}, 
inclination and white dwarf mass in the Wade and Hubeny (1998) library. The values of {\it{i}} are 18, 41, 60, 
75 and $81^{\deg}$. The range of accretion rates covers $-10.5 < \log \dot{M} < -8.0$ in steps of 0.5 in the 
log and five different values of the white dwarf mass, namely, 0.35, 0.55, 0.80, 1.03, and 1.2 M$_{\sun}$. 
The process is streamlined by a routine that compares each observed spectrum with the full 900 models using every 
combination of $i$, \.{M} and M$_{wd}$, and provides the model-computed distance and $\chi^{2}$ value of each model fit.
A good sense of the accuracy of our derived accretion rates for IUE FUV spectra of comparable quality is provided 
by a formal error analysis with contours discussed in Winter and Sion (2003). In general, we estimate that our accretion rates 
from these spectra are accurate to within a factor of 2 to 3.

\section{Accretion rates of SW Sex Nova-likes and Non-SW Sex Nova-Likes} 

\subsection{Accretion Rates from Multiparametric Optimization}

 The Puebla et al.(2007) sample contains ten SW Sex systems for which they derived accretion rates
using a method different from ours, known as multiparametric optimization. Of the ten, three systems, RW Tri, RR Pic and LX Ser are listed as "Possible" SW Sex members
on D.Hoard's Big List while V347 Pup is listed as a "Probable" member. Since these systems are listed as members by Rodriguez-Gil et al.
(2007), then we have retained them in the following rough comparison with non-SW Sex nova-likes. For the latter systems, we have 
V592 Cas, CM Del, KR Aur, IX Vel, UX UMa, V794 Aql, V3885 Sgr, VY Scl, RW Sex, RZ Gru and QU Car. if we take the comparison to be within
approximately the orbital period range of 3 to 4.5 hours, then the average accretion rate of the SW Sex systems is $3.6 \times 10^{-9}$ and non-SW Sex members
have \.{M}$ = 3.0\times10^{-9}$ M$_{\sun}$ yr$^{-1}$.  

If we average the SW SE and non-SW Sex systems without regard to orbital period, then
the 13 SW Sex systems have \.{M}$ = 5.0\times10^{-9}$ M$_{\sun}$ yr$^{-1}$  
compared with 11 non-SW Sex nova-likes with \.{M}$ = 1.0\times10^{-8}$ M$_{\sun}$ yr$^{-1}$  
One must caution that this comparison includes non-SW Sex systems having a mixed bag of distance methods, white dwarf masses and fitting
methods with derived accretion rates from different groups. 

Therefore, to facilitate a more uniform comparison between SW Sex and non-SW Sex nova-likes, we chose to (1) enlarge the sample size of SW Sex
and non-SW Sex nova-likes from the IUE and HST archives; (2) restrict the comparison to the period range of the SW Sex stars and; (3) adopt distances 
determined uniformly with the same method. If we restrict our attention to the range of orbital periods between 3 and 4.5 hours
where most nova-like variables appear to be concentrated, we can directly compare the accretion rates
of non-SW Sex systems in that period range to the SW Sex systems in this same range of orbital period.
For this experiment, the model fitting was carried out uniformly for all the systems, SW Sex and non-SW Sex,
using the distances from the method of Knigge (2006) where, ideally, the P$_{orb}$ versus \.{M} relation, would be expected 
to vary commensurately within this restricted period range.

\subsection{N-L Sample with $3 < P_{orb} < 4.5$, Knigge Distances}

In Table 3, we list the best-fitting parameters of our selected sample of SW Sex and non-SW Sex nova-likes (from Tables 1 and 2)
where the entries by column are (1) the system name, (2) nova-like subclass, (3) white dwarf mass, (4) inclination angle,
(5) best-fitting model distance in pc, (6) $\dot{M}_{\sun}$ yr$^{-1}$),  
and (7) $\chi^{2}$ value.

\begin{deluxetable}{lcccccc}
\tablecaption{Nova-Like System Parameters}
\tablenum{3}
\tablecolumns{8}
\tablewidth{0pc}
\tablehead{
\colhead{System}
&\colhead{Type}
&\colhead{M$_{wd}$ (M$_{\sun}$)}
&\colhead{$\it{i}$($\degr$)}
&\colhead{d$_{model}$ (pc)}
&\colhead{$\dot{M}$ (M$_{\sun}$ yr$^{-1}$}
&\colhead{$\chi^{2}$}
}
\startdata

V442 Oph & SW & 0.4   &   75   &   183   &   1$\times10^{-8}$  &  7.01\\
SW Sex   & SW &   0.4   &   75   &   357   &   3$\times10^{-9}$ & 9.86\\
AH Men   & SW & 0.55 & 75 & 116  & 1$\times10^{-9}$  & 2.44\\ 
HL Aqr   & SW &  0.35& 18  & 213  & 1$\times10^{-9}$  & 4.99\\
WX Ari   & SW & 0.35 & 60  & 468 &  1$\times10^{-9}$ & 4.74\\    
LN UMa   & SW & 0.55  & 41  &  405  & 3$\times10^{-10}$ & 5.67\\
BP Lyn   & SW & 0.35 & 75   & 344 & 1$\times10^{-8}$ & 2.809\\
UU Aqr   & SW & 0.35 & 60 & 208  & 1$\times10^{-9}$  & 6.74\\
LQ Peg   & NSW & 0.55 & 60  & 350 & 1$\times10^{-9}$& 6.38\\  
MV Lyr   & NSW & 0.55 & 75  & 442 & 1$\times10^{-8}$ & 5.86\\ 
TT Ari   & NSW & 0.55  & 41  & 65 & 1$\times10^{-9}$ & 3.32\\
VZ Scl   & NSW & 0.55 & 60  & 566 & 3$\times10^{-10}$ & 7.81\\
BZ Cam   & NSW & 0.35  & 60  & 258 & 3$\times10^{-9}$ & 7.09\\
CM Del   & NSW & 0.35 & 75  & 241 & 1$\times10^{-8}$ & 3.61\\
VY Scl   & NSW & 1.03   & 18   & 337  & 1$\times10^{-9}$  & 4.4\\
\hline
\enddata
\end{deluxetable} 

The best fitting accretion disk model for each system is shown in the accompanying multi-part figures where the systems are displayed
in the same order as they are listed in Tables 1,2, and 3. We display in Fig. 1(a) V442 Oph and (b) SW Sex. While the accretion disk 
is the overwhelmingly dominant contributor in the FUV in nova-like variables during their high states, 
these two systems were among seven flagged by Puebla et al.(2007) as possibly having a significant white dwarf flux contribution. This finding 
led us to combine accretion disk models with white dwarf models in the fitting of V442 Oph and SW Sex. The thick solid line represents their 
combination, the dashed curve represents the accretion disk alone, and the dotted curve represents the white dwarf.
We display in Fig. 2 (a) AH Men), (b) HL Aqr, (c) WX Ari, (d) LN UMa, (e) BP Lyn (f) UU Aqr and in Fig. 3 (a) LQ Peg, (b) MV Lyr, (c) TT Ari,
(d) VZ Scl, (e) BZ Cam, (f) CM Del.  The solid line is the
best fitting accretion disk. 
 
\subsubsection{Comments on Individual Systems}

\begin{itemize}
\item{V442 Oph} - The observation of V442 Oph was made when its V-magnitude was 13.7 as indicated by the FES magnitude 
measured with IUE. This is 1.1 magnitudes fainter than its visual magnitude in the high state and
0.3 magnitudes brighter than its typical low state visual magnitude of 14.0 although it has been observed as faint as
15.5. Given this brighter low state, it may not surprising Puebla et al.(2007) predicted a WD flux contribution.  
With the system inclination and white dwarf mass constrained by the range of values
published in the literature (see Table 1) and the distance of 183 pc, the best-fit (see Fig. 1a) is a combination of 
an accretion disk model with $\it{i} = 75\degr$, M$_{wd}$ = 0.4 M$_{\sun}$ with an accretion rate 
1$\times10^{-8}$ M$_{\sun}$ yr$^{-1}$ and a white dwarf with log $g = 7$, T$_{eff}$ = 23,000K but only a modest improvement 
in the fit over a disk alone. Puebla et al.(2007) derived $3\times10^{-9}$ M$_{\sun}$ yr$^{-1}$ for 130 pc. However, 
Shafter and Szkody (1983) derived $1.0\times10^{-8}$ M$_{\sun}$ yr$^{-1}$ from three SWP spectra which had essentially the same flux level as
SWP14731 and the same IUE FES magnitudes at the time their spectra were taken.

\item{SW Sex} - We co-added a closely spaced time series of
three IUE spectra to improve the signal. Puebla et al.(2007) predicted a significant WD contribution.
The co-added spectrum was best-fit by a combination of an accretion disk with M$_{wd}$ = 0.4 M$_{\sun}$, $\it{i} = 75\degr$, an accretion rate
of $3\times10^{-9}$ M$_{\sun}$ yr$^{-1}$ and a white dwarf with log $g = 7$, T$_{eff}$ = 21,000K for a distance
of 357 pc and $\chi^{2} = 9.86$. This combined fit (see Fig. 1a) agrees with the 
accretion rate value derived by Puebla et al (2007). 

\item{AH Men} - AH Men is normally at V$= 13.2$ but has been observed as faint as V = 14. There are several high state spectra and two low state spectra.
(Mouchet et al.1996) estimated the reddening to be $E(B-V) = 0.12$ and derived an accretion rate of $3\times10^{-9}$ M$_{\sun}$ yr$^{-1}$ 
using standard black body accretion disk models. We modeled one of the highest flux level IUE spectra (SWP41399), displayed in Fig. 2(a).

\item{HL Aqr} - This low inclination object, PHL227, is virtually a twin of V3885 Sgr in both the optical and the FUV (Hunger, Heber and Koester 1985).
The H$\alpha$ line of HL Aqr shows significant blueshifted absorption modulated at the orbital period.  
While the high inclination SW Sex stars are dominated by emission, HL Aqr and other low inclination SW Sex stars show dominant absorption.
Our modeling with the Knigge distance yields an accretion rate of 1$\times10^{-9}$ M$_{\sun}$ yr$^{-1}$.   

\item{WX Ari} - This is the first time that the IUE spectrum and a derived accretion rate has appeared for WX Ari, a definite SW Sex star in D.W. Hoard's
Big List. The inclination is uncertain. Our best-fit favors 60 degrees.

\item{LN UMa}- Listed as a "probable" SW Sex member. This is the first time the IUE spectrum and a derived accretion has been published for LN UMa.
LN UMa was discovered, like other nova-likes, as a thick disk CV in the Palomar-Green Survey.

\item{BP Lyn} - This is first publication of the IUE spectrum. The continuum is very steeply rising toward shorter wavelengths. Many strong 
absorption features are present.

\item{MV Lyr} - The IUE archival spectrum was unfortunately not obtained during a high state of MV Lyra but instead an intermediate brightness state.
Of ten IUE spectra, two were taken during intermediate states and eight during low states. Therefore, we have removed it from the 
non-SW Sex sample compared in the same orbital period range as the SW Sex stars. However, Linnell et al. (2005) modeled an HST spectrum of MV Lyra
taken during a high state. They obtained an accretion rate \.{M}$ = 3.0\times10^{-9}$ M$_{\sun}$ yr$^{-1}$ for a distance of 505 pc, 
well within the Knigge et al. distance range. We have included the accretion rate derived from the HST STIS spectrum of MV Lyra in its high state
in our comparison of SW Sex and non-SW Sex accretion rates. 

\item{TT Ari} - There are numerous high state IUE archival spectra. TT Ari is listed as a "Possible" SW Sex member on D.W.Hoard's Big List. It has 
exhibited positive superhumps during high states and negative superhumps during low states. The white dwarf temperature (39,000K) is well-determined
from HST spectral data taken during a low state. 

\item{VZ Scl} - Eclipsing system with optical spectra out of eclipse revealing strong emission lines of the Balmer series, He II and He I.

\item{BZ Cam} - This nova-like object has highly variable wind outflow, a bow shock nebula, strong, highly variable wind absorption with 
pronounced P Cygni profiles  in C IV, Si IV and very short timescale line profile variations. The origin of the bow shoock nebula remains unclear.

\item{CM Del} - Puebla et al. obtained  \.{M}$ = 4.0\times10^{-9}$ M$_{\sun}$ yr$^{-1}$ but the spectrum they analyzed, SWP15280, was in a low 
to intermediate state with a flux level at 1350A of $5\times 10^{-14}$. They missed the higher brightness state spectrum, SWP14707, which has a flux 
level at 1350\AA\  of $1.2 \times 10^{-13}$ ergs/cm$^2$/s/\AA. This is the spectrum we have modeled.

\item{VY Scl} - The orbital period remains uncertain. Puebla et al (2007) derived an accretion rate  
\end{itemize}

The resulting accretion rates of the two groups of nova-likes are compared in Table 4 where the first three columns on 
the left hand side of the table are the SW Sex system name, second column the orbital period in hours, the third column
the accretion rate in M$_{\sun}$ yr$^{-1}$. On the right hand side of the table, the three columns are (1) the non-SW Sex system name
(2) the orbital period in hours; (3) the accretion rate in M$_{\sun}$ yr$^{-1}$.

\begin{deluxetable}{lccccc}  
\tablecaption{Comparison of Accretion Rates of SW Sex Systems Vs. non-SW Sex Systems}
\tablewidth{0pc}
\tablenum{4}	
\tablecolumns{6}
\tablehead{
\multicolumn{3}{c}{SW Sex Systems}
&\multicolumn{3}{c}{Non-SW Sex Systems}
}
\startdata
System
&P$_{orb}(hrs)$
&$\dot{M}$ (M$_{\sun}$yr$^{-1})$
&System
&$P_{orb}(hrs)$
&$\dot{M}$ (M$_{\sun}$ yr$^{-1}$)\\        
\hline
V442 Oph   &	2.88 & $1.0\times10^{-08}$    &	LQ Peg&	2.9           &$1.0\times10^{-09}$\\
AH Men     &	3.05 &	$1.0\times10^{-09}$	&	MV Lyr$^6$   &	3.19	&$3.0\times10^{-09}$\\
SW Sex     &	3.12 &	$3.2\times10^{-09}$    &	TT Ari &	3.3          & $1.0\times10^{-08}$\\
HL Aqr     &	3.25 &	$1.0\times10^{-09}$    &	V751 Cyg$^3$ &  3.47	& $1.0\times10^{-09}$\\
WX Ari     &	3.34 &  $ 1.0\times10^{-09}$	&	VZ Scl &	3.47        & $5.00\times10^{-09}$\\
DW UMa$^1$ &   3.36 & $1.4\times10^{-8}$     &  V794 Aql   &	3.6          & $3.2\times10^{-10}$\\	
LN UMa     &	3.47 &	$3.2\times10^{-10}$	&     BZ Cam  &	3.68         &$3.2\times10^{-09}$\\
BP Lyn       &	3.67 & $1.0\times10^{-08}$	&	CM Del&	3.88  &	1.00$\times10^{-08}$\\
V380 Oph$^2$ & 3.7  &	$1.0\times10^{-09}$    &     KR Aur$^4$   &   3.90   & $6.9\times10^{-9}$\\
UU Aqr     &   3.93 & $1.0\times10^{-9}$      &   VY Scl&	3.99       &	$1.0\times10^{-09}$\\           
            &       &                       &     IX Vel$^5$      &    4.65  &   $5\times10^{-9}$\\
                                               				\hline
\enddata
\tablerefs{
\\$^1$ DW UMa   (Puebla et al. 2007)\\
$^2$ V380 Oph (Zellem et al. 2009)\\
$^3$ V751 Cyg (Zellem et al. 2009)\\
$^4$ KR Aur   (Mizusawa et al.2009)\\
$^5$ IX Vel   (Linnell et al. 2007)\\
$^6$ MV Lyr   (Linnell et al. 2005)\\
}

\end{deluxetable}                                   
                   
First, we computed the average accretion rates of the 16 SW Sex systems and non-SW Sex systems in this paper within approximately the same orbital 
period range of 3 to 4.5 hours, the period range where 27 out 35 SW Sex stars reside (Rodriguez-Gil et al. 2007). The average accretion
for the SW Sex systems is 2.7$\times10^{-09}$ (M$_{\sun}$ yr$^{-1})$ and for the non-SW Sex systems 4.2$\times10^{-09}$ (M$_{\sun}$ yr$^{-1})$.
Within the uncertainty of our accretion rates, there is no difference in the accretion rates of the two groups. Second, we added five
additional nova-likes, two SW Sex systems and three non-SW Sex systems discussed elsewhere but whose accretion rates were determined with the same
model grid as used in this paper and with Knigge (2006) distances. This amounted to 10 SW Sex systems and eleven non-SW Sex systems. Once again there 
is virtually no difference in the average accretion rates. In view of these results, we do not believe that SW Sex have higher
secular accretion rates that other CVs, specifically the nova-like systems that do not exhibit the 
SW Sex spectroscopic characteristics and behavior.

\section{Conclusions}

(1) We have examined the accretion rates of nova-like systems of the SW Sex and non-SW Sex subclasses that were derived by the 
multi-parametric optimization method of Puebla et al.(2007). If the two subclasses are compared in the same orbital period range of
3 to 4.5 hours, then the average accretion rates of the two subclasses are essentially the same, 3.6$\times10^{-9}$ M$_{\sun}$ yr$^{-1}$ 
for the SW Sex systems and 3.0$\times10^{-9}$ M$_{\sun}$ yr$^{-1}$ for the non-SW Sex systems.  
If the average accretion rates of the two groups are computed with no restriction on the orbital periods of the two groups, then
the SW Sex systems have \.{M}$ = 5\times10^{-9}$ M$_{\sun}$ yr$^{-1}$ and the non-SW Sex systems have 
\.{M}$ = 1.0\times10^{-8}$ M$_{\sun}$ yr$^{-1}$. Using a different approach, we enlarged the sample of 
SW Sex and non-SW Sex stars, restricted attention to the orbital period range of 3 to 4.5 hour, used distances from the 
method of Knigge (2006) and applied a different methodology for determining accretion rates.
We find that the non-SW Sex systems have an average accretion rate $\dot{M}$ =
3.4$\times10^{-9}$ M$_{\sun}$ yr$^{-1}$ and the SW Sex systems in the sample also have 3.4$\times10^{-9}$ M$_{\sun}$ yr$^{-1}$.
Therefore, based upon two independent methods of deriving accretion rates, that of Puebla et al. (2007) and the approach in this paper
with Knigge (2006) distances, we find little evidence to support the suggestion that the SW Sex systems have higher than average accretion rates among
the nova-like systems, a possibility raised by Rodriguez-Gil et al (2007). Therefore, it is likely that the SW Sextantis 
phenomenon, particularly their high optical luminosities, must be attributed to some other factor or characteristic of the 
systems than higher than average accretion rates. Among these possibilities are magnetic accretion and nuclear burning 
(Honeycutt 2001; Honeycutt and Kafka 2004). However, this conclusion applies only to the average accretion rates of the two
groups. This does not rule out the possibility that the SW Sex phenomena is exhibited when the accretion flow in a given system
has undergone a temporary large increase at the time an observation of a nova-like is obtained thus leading to the object being
observed as an SW Sex star (Groot et al.2004). We see little to suggest that SW Sex stars have higher than average accretion rates than other nova-likes
in the same range of orbital period.

(2) Given the high average value of \.{M} in
nova-likes and the fact that their high states 
are generally longer in duration than their low states, the rate of accretion onto the underlying
white dwarf and hence a higher degree of compressional heating would be expected. Thus, surface temperatures of nova-likes 
should be higher than in dwarf novae at the same orbital period.
This should be true if one accepts a correlation between CV orbital period and \.{M} such as shown by Patterson (1984).
There is some preliminary evidence that this is the case when one 
compares the surface temperatures of white dwarfs in nova-like variables to the white dwarfs in dwarf 
novae (Hamilton and Sion 2007; Godon et al. 2008). However, these studies have relied upon the relatively rare situation when the
nova-like drops into a deep low state and the white dwarf is exposed to FUV spectroscopic
observation. We point out that a number of SW Sex systems have lower orbital inclinations. In these systems, the upper hemisphere
of the underlying white dwarf and the inner boundary layer would not be obscured.   
It would be interesting to find evidence of cooler white dwarfs in SW Sex systems and non-SW Sex systems, unlike
the typically high temperatures found for DW UMa and MV Lyr, since such cooler temperatures
would be unexpectedly low for the derived high rates of time-averaged mass transfer.
Unfortunately, we are unable to reliably characterize the white dwarf temperatures in the lower inclination SW Sex systems
due to: (1) the poor quality FUV spectra; (2) the overwhelming luminosity of the bright (high state) accretion disk
and (3) the lack of FUV spectra down to the Lyman limit (e.g. FUSE) where the flux contribution of a bright accretion disk 
and hot white dwarf photosphere can be more easily disentangled.

Finally, given the small sample size of exposed white dwarfs in nova-like systems, it is 
particularly important to catch more nova-like systems in their low states for both 
ground-based optical and space observations.

\acknowledgments 

This research utilized the Big List of SW Sextantis Stars, which is maintained by D. W. Hoard.
This work was supported by NSF grant AST0807892 to Villanova University. 
Some or all of the data presented in this paper were obtained from the Multi mission Archive at the Space 
Telescope Science Institute (MAST). STScI is operated by the Association of Universities for Research in 
Astronomy, Inc., under NASA contract NAS5-26555. Support for MAST for non-HST data is provided by the NASA 
Office of Space Science via grant NAG5-7584 and by other grants and contracts.

\section{Figure Captions}

\begin{figure}
\plotone{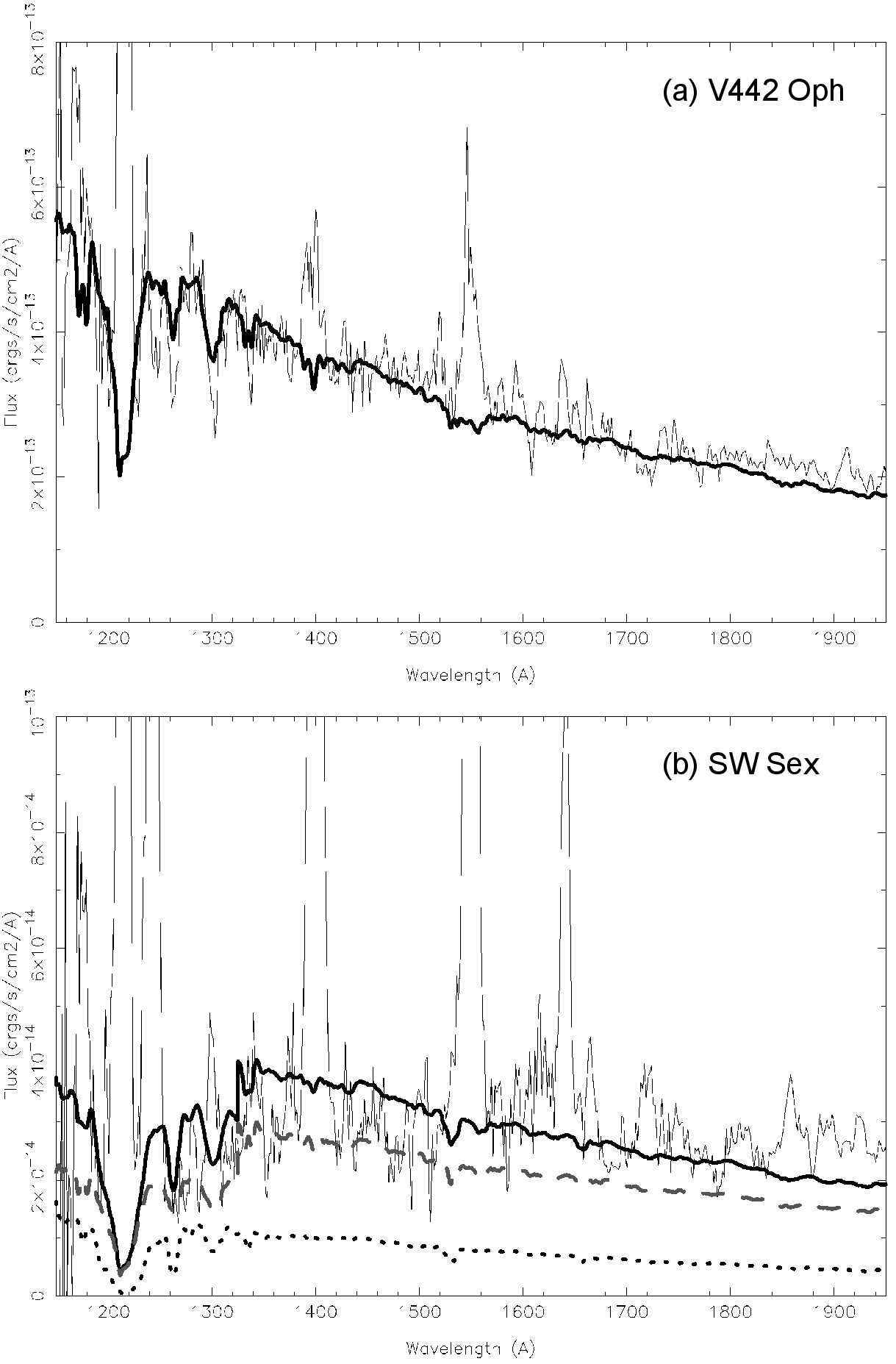}
\caption{Flux density versus wavelength plots of IUE 
archival SWP spectra for (a) V442 Oph, (b) SW Sex. 
The WD model flux is denoted by a dotted
line, the accretion disk model flux alone by a dashed line and the combined flux by the
solid line.}
\end{figure}

\begin{figure}
\plotone{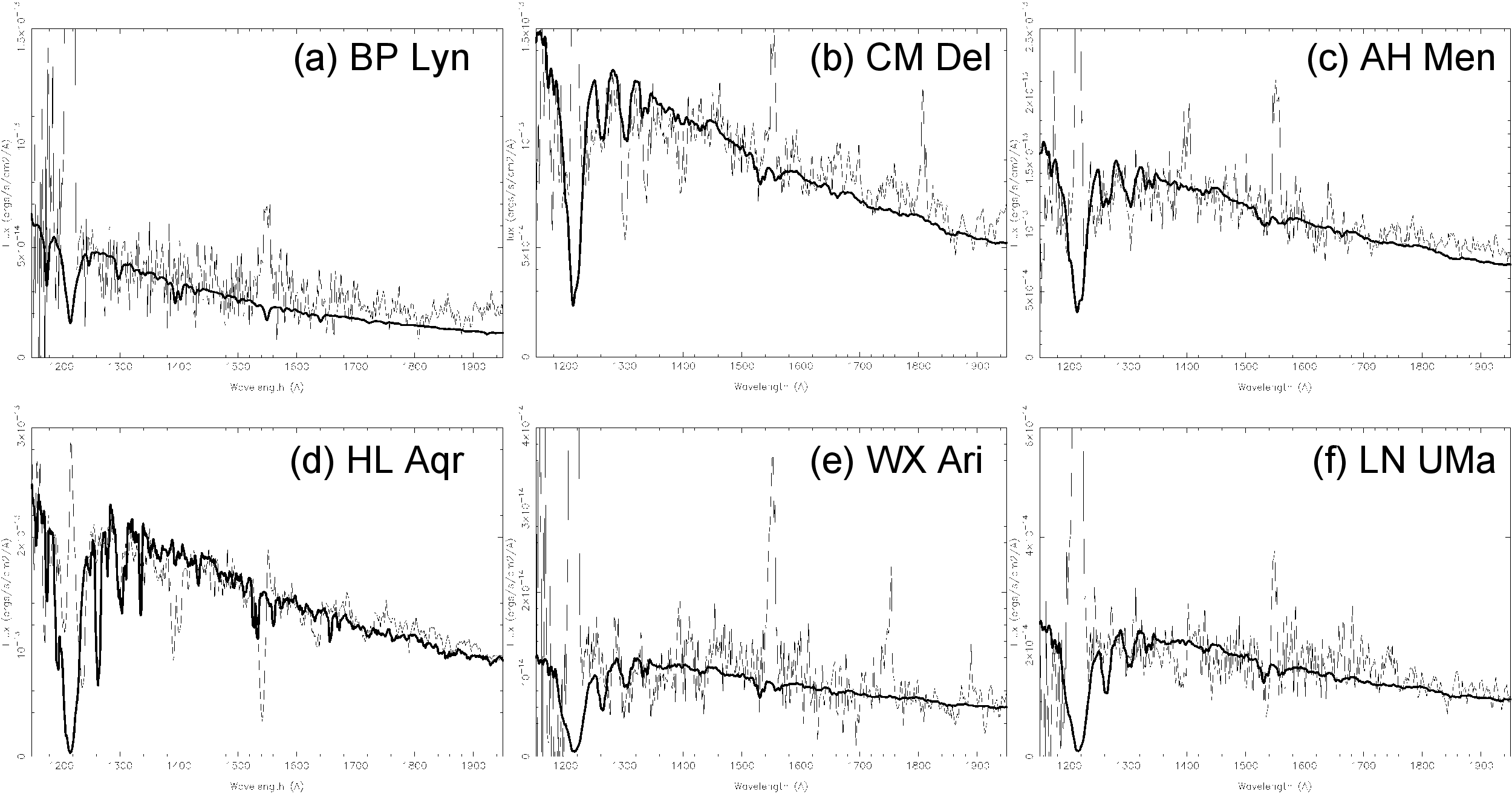}
\caption{Flux density versus wavelength plots of IUE archival SWP spectra for (a) BP Lyn, (b) CM Del, 
(c) AH Men, (d) HL Aqr, (e) WX Ari, (f) LN UMa. The best fitting accretion disk model is
shown with the thick solid line.}
\end{figure}

\begin{figure}
\plotone{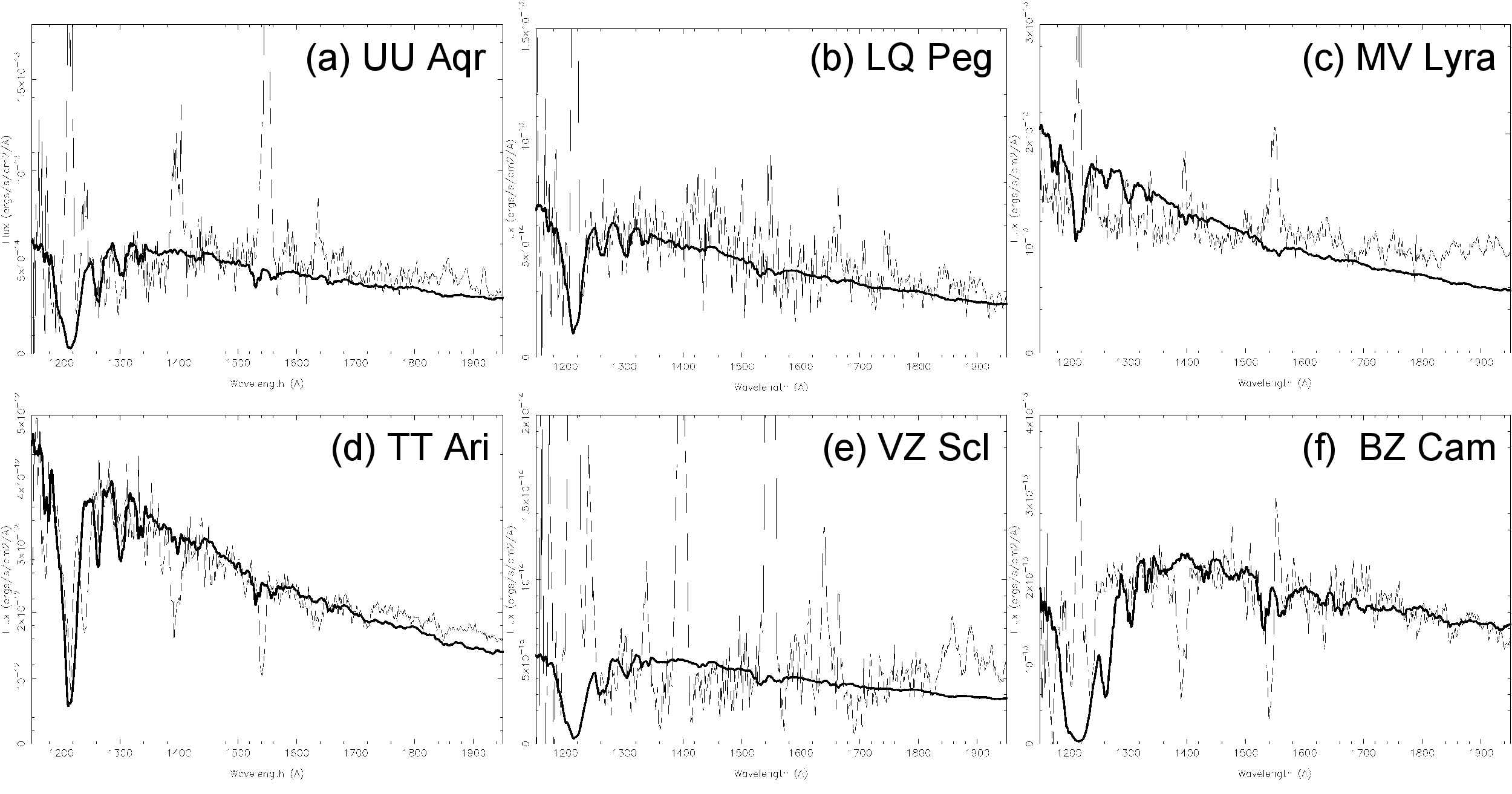}
\caption{Same as Figure 2 except for (a) UU Aqr, (b) LQ Peg, (c) MV Lyr,
(d) TT Ari, (e) VZ Scl, (f) BZ Cam.}
\end{figure}


\begin{thebibliography}{}

\bibitem[]{}
Ballouz, R.-L., Sion, E.M., Gaensicke, B., Long, K. 2009, ApJ, in preparation

\bibitem[]{}
Bruch, A., \& Engel, A. 1994, A\&A Suppl.Ser., 104, 79

\bibitem[]{}
Diaz et al. 1996, ApJ, 459, 236

\bibitem[]{}
Frank, J., King, A.R., \& Raine, D.J. 1992, Accretion Power in Astrophysics (Cambridge: Cambridge Univ. Press)

\bibitem[]{}
Godon et al. 2007, ApJ, 656, 1103

\bibitem[]{}
Godon et al. 2008, ApJ, 679, 1447

\bibitem[]{}
Godon, P., Sion, E., Barrett, P., Szkody, P., \& Schlegel, E.2008, ApJ, 687, 532

\bibitem[]{}
Groot, P., Rutten, R.G.M., \& van Paradijs, J.2004, A\&A, 417, 283

\bibitem[]{}
Hamilton, Ryan T., Sion, E. M., 2008, PASP, 120, 165

\bibitem[]{}
Hamilton, Ryan T., Urban, Joel A.; Sion, Edward M.; Riedel, Adric R.; Voyer, Elysse N.; Marcy, John T.; Lakatos, Sarah L.2007, ApJ, 667, 1139

\bibitem[]{}
Hoard et al.2003, AJ, 126, 2473

\bibitem[]{}
Honeycutt, K.2001, PASP, 113, 473

\bibitem[]{}
Honeycutt, R. K.,\& Kafka, S.,2004, AJ, 128, 1279

\bibitem[]{}
Hubeny, I. 1988, Comput. Phys. Comun., 52, 103

\bibitem[]{}
Hubeny, I., and Lanz, T. 1995, ApJ, 439, 875

\bibitem{}{}
Hunger, K., Heber, U., \& Koester, D.1985, A\&A, 149, 4

\bibitem[]{}
Knigge, C.2006, MNRAS, 373, 484

\bibitem[]{}
LaDous, C. 1991, A\&A, 252, 100

\bibitem[]{}
Livio, M., \& Pringle, J.E. 1998, ApJ, 505, 339

\bibitem[]{}
Linnell, A. et al.2005, ApJ, 624, 923

\bibitem[]{}
Linnell, A. 2007, ApJ, 662, 1204

\bibitem[]{} 
Mizusawa, T., et al.2009, PASP, in preparatioon 

\bibitem[]{}
Mouchet, M. et al. 1996, A\& A, 306, 212

\bibitem[]{}
Patterson, J.1984, ApJS, 54, 443

\bibitem[]{}
Puebla, R.E., Diaz, M.P., \& Hubeny, I.2007, AJ, 134, 1923 

\bibitem[]{}
Ritter, H., \& Kolb, U. 2003, A\&A, 404, 301(update RKcat7.10)

\bibitem[]{}
Rodriguez-Gil, P., Schmidtobreick, L., Gaensicke, B.T. 2007, MNRAS 374, 1359

\bibitem[]{}
P. Rodriguez-Gil, B. T. G¨ansicke, H.-J. Hagen, S. Araujo-Betancor,
A. Aungwerojwit, C. Allende Prieto, D. Boyd, J. Casares, D. Engels,
O. Giannakis, E. T. Harlaftis, J. Kube, H. Lehto, I. G. Martinez-Pais,
R. Schwarz, W. Skidmore, A. Staude, and M. A. P. Torres, 2007, MNRAS, 377, 1747

\bibitem[]{}
Shafter,A. W., Wheeler,J. C., \& Cannizzo, J. K., 1986, ApJ, 305, 261

\bibitem[]{}
Verbunt, F.1987, A\&A, 71, 339

\bibitem[]{}
Wade, R.A. \& Hubeny, I. 1998, ApJ, 509, 350.

\bibitem[]{}
Warner, B. 1995, Cataclysmic Variable Stars (Cambridge: Cambridge Univ. Press)

\bibitem[]{}
Warner, B. 2008, private communication

\bibitem[]{}
Winter, L., \& Sion, E.M.2003, ApJ, 582, 352

\bibitem[]{}
Wood, M. 1995, in White Dwarfs, Proceedings of the 9th European Workshop on White
Dwarfs Held at Kiel, Germany, 29 August - 1 September 1994. Lecture Notes in
Physics, Vol.443, edited by Detlev Koester and Klaus Werner. Springer-Verlag, Berlin
Heidelberg New York, 1995., p.41.
 
\bibitem[]{}
Wu, K., Warner, B., Wickramasinghe, D.T. 1995, Proc.Ast.Soc.Austral.12:1, p.60 

\bibitem[]{}
Zellem, R., Hollon, N., Ballouz, R.-L., Sion, E.M., Gaensicke, B.2009, submitted to PASP.

 \end{thebibliography}
\end{document}